\documentclass{article}
\usepackage{amsmath,amsthm,amssymb}
\usepackage{array, longtable}
\newtheorem{Theorem}{Theorem}[section]
\newtheorem{Lemma}[Theorem]{Lemma}
\newtheorem{Remark}[Theorem]{Remark}
\newtheorem{Definition}[Theorem]{Definition}
\newtheorem{Corollary}[Theorem]{Corollary}
\newtheorem{Proposition}[Theorem]{Proposition}
\newtheorem{Example}[Theorem]{Example}

\numberwithin{equation}{section}
\numberwithin{table}{section}

\def\Z{{\Bbb Z}}

\def\pf{\par{\bf Proof.}~ }

\def\M{{\varphi}}  
\def\CRT{\,\mathop{=}\limits^{\mbox{\rm\tiny CRT}}\,}

\begin{document}

\title{%Even-like Duadic Constacyclic Codes
Iso-Orthogonality and Type-II Duadic Constacyclic Codes}

\insert\footins{\footnotesize
{\it Email addresses}:  yfan@mail.ccnu.edu.cn (Y. Fan);
554701169@qq.com (L. Zhang).}

\author{Yun Fan\quad and\quad Liang Zhang\\
\small Dept of Mathematics,
\small  Central China Normal University, Wuhan 430079, China}

\date{}

\maketitle

\begin{abstract}

Generalizing even-like duadic cyclic codes and Type-II duadic
negacyclic codes, we introduce even-like (i.e.,Type-II) and
odd-like duadic constacyclic codes,
and study their properties and existence.
We show that even-like duadic constacyclic codes are isometrically orthogonal,
and the duals of even-like duadic constacyclic codes
are odd-like duadic constacyclic codes.
We exhibit necessary and sufficient conditions for the existence
of even-like duadic constacyclic codes.
A class of even-like duadic constacyclic codes 
which are alternant MDS-codes is constructed.


\end{abstract}

\section{Introduction}

The study of duadic cyclic codes was initiated by
Leon, Masley and Pless \cite{LMP}, and
attracted many attentions,  e.g. \cite{P87, S, DP,DLX,HK}.
The research on duadic cyclic codes over finite fields has greatly developed,
see \cite[Ch.6]{HP}.
Rushanan \cite{R} generalized duadic cyclic codes to 
the duadic group codes, and many results about the existence of
such codes, especially the duadic abelian codes, were obtained, e.g. \cite{WZ, AKS}.

Note that most of the studies on duadic cyclic codes over finite fields
consider the semisimple case, i.e.,
the length of the codes is coprime to the cardinality of the finite field.
In that case, duadic cyclic codes are not self-dual;
the key obstruction is the $1$-dimensional cyclic code with
check polynomial $X-1$, which is invariant by any multipliers.
By appending with one bit, self-dual extended cyclic codes might be obtained.
It is the same for group codes, more generally, for transitive permutation codes,
see \cite{FZ}.

Another perspective to carry the research forward is to consider constacyclic codes.
Let $F_q$ be the finite field with $q$ elements, where $q$ is a power of a prime,
let $F_q^*$ be the multiplicative group consisting of the non-zero elements of $F_q$.
Let $n$ be a positive integer coprime to $q$,
and $\lambda\in F_q^*$ with $r={\rm ord}_{F_q^*}(\lambda)$, where
${\rm ord}_{F_q^*}(\lambda)$ denotes the order of $\lambda$
in the group $F_q^*$.
Any ideal $C$ of the quotient algebra $F_q[X]/\langle X^n-\lambda\rangle$
is called a {\em $\lambda$-constacyclic code} of length $n$ over $F_q$,
where $F_q[X]$ denotes the polynomial algebra over $F_q$ and
$\langle X^n-\lambda\rangle$ denotes the ideal generated by $X^n-\lambda$.
In the following we always use the three numbers $q$, $n$, $r$
to parametrize the $\lambda$-constacyclic code $C$.
If $r=1$ (i.e. $\lambda=1$) then $C$ is just a cyclic code.
If $r=2$ (i.e. $\lambda=-1$) then $C$ is named {\em negacyclic code}.

Aydin {\it et al} \cite{ASR} exhibited the BCH bound of constacyclic codes.
Dinh {\it et al} \cite{DL,D} studied
constacyclic codes and showed that self-duality happens for
 (and only for) negacyclic codes.

Blackford \cite{Bl08, Bl13} contributed very much to
the study of the duadic constacyclic codes.
Let $\Z_{nr}$ be the residue ring of the integer ring $\Z$ modulo $nr$.
The set of roots of the polynomial $X^n-\lambda$ in its splitting field
corresponds to a subset  of $\Z_{nr}$:
$1+r\Z_{nr}=\{1,1+r,\cdots,1+r(n-1)\}$.
The multipliers act on this set $1+r\Z_{nr}$.
In this way, Blackford~\cite{Bl08} obtained all self-dual negacyclic codes,
introduced Type-I and Type-II %(even-like) 
duadic splittings in the negacyclic case (i.e. $r=2$),
and proved the existence of Type-II duadic negacyclic codes
for the case when $n$ is even but $n/2$ is odd.
Further, in \cite{Bl13},
it was shown that the Type-I duadic constacyclic codes
are just the so-called {\em iso-dual} constacyclic codes.

Type-I polyadic (including duadic) constacyclic codes
were studied in \cite{CDFL}. In terms of the Chinese Remainder Theorem, the set
$1+r\Z_{nr}$ and the multiplier group can be decomposed suitably.
Necessary and sufficient conditions for the existence of such codes
were obtained. Some generalized Reed-Solomon or alternant constacyclic codes
were constructed from Type-I polyadic constacyclic codes.

In this paper, generalizing even-like (Type-II) and odd-like duadic negacyclic codes,
we introduce even-like (i.e.,Type-II) and odd-like duadic constacyclic codes,
and study their properties and existence.

In Section 2, necessary notations and fundamentals are described.

In Section 3,
with isometries between constacyclic codes, 
even-like (i.e., Type-II) and odd-like duadic constacyclic codes are defined, and
a relationship between the two kinds of duadic constacyclic codes are
exhibited (Theorem \ref{even odd} below).
As known, even-like duadic constacyclic codes are not self-orthogonal in general.
We show that they are {\em iso-orthogonal} and, up to some sense,
they are the maximal iso-orthogonal pairs of constacyclic codes
(Theorem \ref{iso-orth}). 

For the existence of Type-I duadic constacyclic codes,
a necessary and sufficient condition has been obtained in \cite[Th.4]{CDFL},
see Lemma \ref{group n_r}(iii) below also.
In Section 4, we present necessary and sufficient conditions for
the existence of Type-II duadic constacyclic codes,
see Theorem \ref{existence} below, where the cyclic case and the negacyclic case
are included as straightforward consequences.

In Section 5, a class of alternant MDS-codes is constructed from
even-like duadic constacyclic codes (Proposition \ref{a GRS}),
and some specific examples are presented.

\section{Preliminaries}

Throughout this paper, $F_q$ is the finite field with $q$ elements, 
$\lambda \in F_q^*$ has multiplicative order $r$, 
and $n$ is a positive integer that is relatively prime to $q$.
Note that the order of the multiplicative group $|F_q^*|=q-1$, 
hence $r\,|\, (q-1)$. Following \cite{Bl13}, 
we abbreviate the quotient algebra by
\begin{equation}\label{R_n,lambda}
 R_{n,\lambda}=F_q[X]/\langle X^n-\lambda\rangle.
\end{equation}
If $C$ is an ideal of $R_{n,\lambda}$ (i.e., a
$\lambda$-constacyclic code $C$ over $F_q$ of length $n$),
we say that $C\subseteq R_{n,\lambda}$ is a $\lambda$-constacyclic code. 

In this paper we always assume that $\theta$ is a primitive $n r$-th root of unity
(in a suitable extension of $F_q$) such that $\theta^n=\lambda$.
As mentioned in the Introduction, 
the set of roots of $X^n-\lambda$ corresponds to the
subset $P_{n,\lambda}$ of the residue ring $\Z_{nr}$, which is defined by:
\begin{equation}\label{P_n,lambda}
P_{n,\lambda}=1+r\Z_{nr}
=\{1+rk~({\rm mod}~nr)\mid k\in\Z_{nr}\},
\end{equation}
so that
$$
X^n-\lambda=\prod_{i\in P_{n,\lambda}}(X-\theta^i).
$$

By $\Z_{nr}^*$ we denote the multiplicative group
consisting of units of $\Z_{nr}$. 
The group $\Z_{nr}^*$ acts on $\Z_{nr}$ by multiplication.
Precisely, any $t\in\Z_{nr}^*$ induces a permutation $\mu_t$
of the set $\Z_{nr}$ as follows: $\mu_t(k)=tk$ for all $k\in \Z_{nr}$.
Any $\mu_t$-orbit on $\Z_{nr}$ is abbreviated as a {\em $t$-orbit}.
The set of $t$-orbits on $\Z_{nr}$ (i.e., the quotient set by $\mu_t$)
is denoted by $\Z_{nr}/\mu_t$.
For any subset $P\subseteq \Z_{nr}$, the permutation $\mu_t$
transforms $P$ to the subset
$tP=\{tk~({\rm mod}~rn)\mid k\in P\}$.
We say that $P$ is $\mu_t$-invariant if $tP=P$.
If $t\equiv t'~({\rm mod}~r)$ with $1\le t'<r$
(recall that $t$ is coprime to $r$), it is easy to see that
$$tP_{n,\lambda}
=t+rt\Z_{nr}=t+r\Z_{nr}=t'+r\Z_{nr}=t'P_{n,\lambda},$$
and $\theta^j$ for $j\in t'+r\Z_{nr}$ are the roots of
$X^n-\lambda^t=X^n-\lambda^{t'}$.
So we denote $t+r\Z_{nr}=P_{n,\lambda^t}$. 
With this notation, for any $t\in\Z_{nr}^*$ we have
\begin{equation}\label{X^n-lambda^t}
 X^n-\lambda^t=\prod_{i\in P_{n,\lambda^t}}(X-\theta^i),\qquad
 \mbox{where}~~P_{n,\lambda^t}=tP_{n,\lambda}.
\end{equation}
Further, for any $s\in\Z_{nr}^*$, it is easy to see that
$sP_{n,\lambda^t}=P_{n,\lambda^t}$
if and only if $s\in 1+r\Z_{nr}$, i.e., $s\in\Z_{nr}^*\cap(1+r\Z_{nr})$.
We denote
\begin{equation}\label{G_n,r}
G_{n,r}=\Z_{nr}^*\cap(1+r\Z_{nr}),
\end{equation}
which is a subgroup of the group $\Z_{nr}^*$.
We call $G_{n,r}$ the {\em multiplier group}.

Since $r|(q-1)$,  we see that $q\in G_{n,r}$.
The $q$-orbits on  $\Z_{nr}$ are also named
{\em $q$-cyclotomic cosets} (abbreviated to {\em $q$-cosets}) in literature,
so we call them by $q$-cosets.
Obviously, $P_{n,\lambda^t}=t+r\Z_{nr}$
is $\mu_q$-invariant for any $t\in\Z_{nr}^*$. 
For any $q$-coset $Q\in P_{n,\lambda^t}/\mu_q$,
the polynomial $f_Q(X)=\prod_{i\in Q}(X-\theta^i)$ is irreducible
in $F_q[X]$. Thus
\begin{equation}\label{X^n-}
X^n-\lambda^t=\prod_{Q\in P_{n,\lambda^t}/\mu_q}f_{Q}(X)
\end{equation}
is the monic irreducible decomposition in $F_q[X]$.

For any $\mu_q$-invariant subset $P$ of $P_{n,\lambda}$,
We have a polynomial
$$f_P(X)=\prod\limits_{Q\in P/\mu_q}f_Q(X)\in F_q[X].$$
Let $\overline P=P_{n,\lambda}\backslash P$
(which denotes the difference set), i.e.,
$\overline P$ is the complement of $P$ in $P_{n,\lambda}$.
By \eqref{X^n-},
\begin{equation}\label{X^n-lambda}
 f_P(X)f_{\overline P}(X)=X^n-\lambda.
\end{equation}

\begin{Remark}\label{C_P}\rm It is well-known that
for any $\lambda$-constacyclic code $C\subseteq R_{n,\lambda}$ 
there is exactly one $\mu_q$-invariant subset $P\subseteq P_{n,\lambda}$ such that, 
for any $a(X)\in R_{n,\lambda}$,
\begin{itemize}
\item $a(X)\in C$ if and only if
$a(X)f_P(X)\equiv 0~({\rm mod}~X^n-\lambda)$;
\item $a(X)\in C$ if and only if $f_{\overline P}(X)\,\big|\,a(X)$.
\end{itemize}
The polynomial $f_P(X)$ is said to be a {\em check polynomial}
of the $\lambda$-constacyclic code $C$, while
the polynomial $f_{\overline P}(X)$ is said to be
a {\em generator polynomial} of $C$.
In that case we denote $C=C_P$ and call it the $\lambda$-constacyclic code
with {\em check set} $P$ and {\em defining set} $\overline P$
(which corresponds to the zeros of $C_P$). It is easy to see that
\begin{equation}\label{check set}
C_P\subseteq C_{P'} \iff P\subseteq P',\quad
\mbox{for $\mu_q$-invariant subsets $P$, $P'\subseteq P_{n,\lambda}$;}
\end{equation}
which implies that

\hangindent25pt
$\bullet$ {\it mapping a $\lambda$-constacyclic code of length $n$ over $F_q$
to its check set is an isomorphism from
the lattice of $\lambda$-constacyclic codes of length $n$ over $F_q$ onto
the lattice of $\mu_q$-invariant subsets of $P_{n,\lambda}$.}
\end{Remark}

Any element of $R_{n,\lambda}$ has a unique representative:
$a(X)=a_0+a_1X+\cdots+a_{n-1}X^{n-1}$.
We always associate any word
${\bf a}=(a_0,a_1,\cdots,a_{n-1})\in F_q^n$ with
the element $a(X)=a_0+a_1X+\cdots+a_{n-1}X^{n-1}\in R_{n,\lambda}$,
and {\it vice versa}. 
For any $a(X), b(X)\in R_{n,\lambda}$ associated to words 
${\bf a}=(a_0,a_1,\cdots,a_{n-1}),{\bf b}=(b_0,b_1,\cdots,b_{n-1})\in F_q^n$,
the Hamming weight ${\rm w}(a(X))$ 
is defined by the Hamming weight of the word ${\bf a}$; 
the Euclidean inner product of $a(X)$ and $b(X)$ is defined by
the Euclidean inner product of the words ${\bf a}$ and ${\bf b}$:
\begin{equation}\label{inner}
 \big\langle a(X), b(X)\big\rangle=
 \big\langle {\bf a},{\bf b}\big\rangle=\sum_{i=0}^{n-1}a_ib_i.
\end{equation}
For $C\subseteq F_q^n$, denote
$$C^{\bot}=\big\{ {\bf a}\in F_q^n\,\big|\,
 \big\langle {\bf c}, {\bf a}\big\rangle=0,~\forall~{\bf c}\in C\big\},
$$
which is called the {\em (Euclidean) dual code} of $C$.
It is known that, for a $\lambda$-constacyclic code~$C$,
the dual code $C^{\bot}$ is in fact a $\lambda^{-1}$-constacyclic code,
see \cite{Bl08, D}, or see Lemma~\ref{dual code} below for more precise description.

\begin{Remark}\label{group theory}\rm
Let $\Gamma$ be a finite set, and $\sigma$ be a permutation of $\Gamma$.
We need the following group-theoretical results (cf. \cite[Lemmas 6-8]{CDFL}).
For fundamentals about groups, please refer to \cite{AB}.
\begin{itemize}
\item[(i)]
%Assume that $\sigma$ is a permutation of a finite set $\Gamma$.
The group generated by $\sigma$ is denoted by $\langle\sigma\rangle$.
The orbits of $\langle\sigma\rangle$ on $\Gamma$ is abbreviated by
{\em $\sigma$-orbits}. The length of the $\sigma$-orbit containing
$k\in\Gamma$ is equal to the index
$|\langle\sigma\rangle:\langle\sigma\rangle_k|$, where
$\langle\sigma\rangle_k$ denotes the subgroup consisting of
the elements of $\langle\sigma\rangle$ which fix $k$.
In particular, the length of any $\sigma$-orbit is a divisor of the
order ${\rm ord}(\sigma)$ of $\sigma$.
\item[(ii)]
There is a  partition $\Gamma=\Gamma_1\cup\Gamma_{2}$ such that
$\sigma(\Gamma_1)=\Gamma_{2}$ and $\sigma(\Gamma_2)=\Gamma_{1}$
if and only if the length of every $\sigma$-orbit on $\Gamma$ is even.
\item[(iii)]
Further assume that $\sigma'$ is a permutation of a finite set $\Gamma'$,
hence $(\sigma,\sigma')$ is a permutation of the product
$\Gamma\times\Gamma'$.  Then the order of the permutation $(\sigma,\sigma')$
is equal to the least common multiple of the order of $\sigma$ and 
the order of $\sigma'$; the length of the $(\sigma,\sigma')$-orbit
containing $(k,k')\in\Gamma\times\Gamma'$ is equal to the
least common multiple of the length of the $\sigma$-orbit containing $k\in\Gamma$
and the length of the $\sigma'$-orbit containing $k'\in\Gamma'$.
\item[(iv)]
Assume that a finite group $G$ acts on a finite set $\Gamma$
and $N$ is a normal subgroup of $G$.
Let $\Gamma/N$ be the set of $N$-orbits on $\Gamma$,
called the {\em quotient set} of $\Gamma$ by $N$.
Then the quotient group $G/N$ acts on the quotient set $\Gamma/N$.
In particular, for $\sigma\in G$, the length of any $\sigma$-orbit
on the quotient set $\Gamma/N$ is a divisor of the order
of $\sigma$ in the quotient group $G/N$.
\end{itemize}
\end{Remark}

\section{Three kinds of duadic constacyclic codes}

We keep notations introduced in Section 2.
In this section we define three kinds of duadic constacyclic codes and
study their properties. We begin with a class of isometries 
between constacyclic codes, which is a generalization of the multipliers
for cyclic codes (cf. \cite[\S4.3, eq.(4.4)]{HP}).

\begin{Lemma}\label{isometry}
Let $t$ be an integer coprime to $nr$ and $\bar t$ be a positive integer
such that $t\bar t=1~({\rm mod}~nr)$.
Then the following map
(where $R_{n,\lambda^t}=F_q[X]/\langle X^n-\lambda^t\rangle$,
cf. Eq.~\eqref{R_n,lambda})
$$
\M_t:  R_{n,\lambda} \longrightarrow  R_{n,\lambda^t},\quad
  \sum\limits_{i=0}^{n-1}a_iX^i \longmapsto
\sum\limits_{i=0}^{n-1}a_i X^{\bar t i}
 ~({\rm mod}~{X^n-\lambda^{t}}),
$$
is an algebra isomorphism and preserves the Hamming weights of words, i.e.,
${\rm w}(\M _t(a(X)))={\rm w}(a(X))$ for any $a(X)\in R_{n,\lambda}$.
\end{Lemma}

\pf 
For the polynomial algebra $F_q[X]$ 
%(i.e., the negative powers of $X$ are allowed), 
it is obvious that the following map
$$
F_q[X]  \longrightarrow  F_q[X],~~
\sum\limits_{i}a_iX^i\longmapsto
\sum\limits_{i}a_i X^{\bar t i},
$$
is an algebra homomorphism. Hence it induces an algebra homomorphism:
$$\begin{array}{crcl}
\hat\M _t: & F_q[X] & \longrightarrow
  & F_q[X]/\langle X^n-\lambda^{t}\rangle,\\[3pt]
&\sum\limits_{i} a_i X^i &\longmapsto&
\sum\limits_{i}a_i X^{\bar t i}
\pmod{X^n-\lambda^{t}}.
\end{array}$$
In the algebra $R_{n,\lambda^t}=F_q[X]/\langle X^n-\lambda^{t}\rangle$
we have the following computation:
$$
\hat\M _{t}(X^n-\lambda)=X^{n\bar t}-\lambda
=(\lambda^{t})^{\bar t}-\lambda=0 \pmod{X^n-\lambda^{t}}.
$$
The algebra homomorphism $\hat\M _t$ induces an
algebra homomorphism as follows.
$$\begin{array}{crcl}
\M _t: & F_q[X]/\langle X^n-\lambda\rangle & \longrightarrow
  & F_q[X]/\langle X^n-\lambda^{t}\rangle,\\[3pt]
&\sum\limits_{i=0}^{n-1}a_iX^i&\longmapsto&
\sum\limits_{i=0}^{n-1}a_i X^{\bar t i} \pmod{X^n-\lambda^t}.
\end{array}$$
That is, $\M_t(a(X))=a(X^{\bar t})\!\pmod{X^n-\lambda^t}$ 
for $a(X)\in R_{n,\lambda}$.
Since $\lambda^r=1$, the $\lambda^t$ and the algebra homomorphism  
$\M_t$ are uniquely determined by $t$ (independent of the choice of $\bar t$)
up to modulo $nr$.

For any $i$ with $0\le i<n$,
there are a unique $t_{i}$ with $0\le t_i<n$
and a unique $q_i$ such that $\bar t i=nq_i+t_i$.
Thus, in $R_{n,\lambda^t}$ we have
\begin{equation}\label{psi_s}
\M_t\Big(\sum_{i=0}^{n-1}a_iX^i\Big)=\sum_{i=0}^{n-1}a_iX^{\bar t i}
=\sum_{i=0}^{n-1}a_i\lambda^{t q_i}X^{t_i}.
\end{equation}
The map $i\mapsto t_i$ is a permutation of the index set $\{0,1,\cdots,n-1\}$
and all ${\lambda^{t q_i}\ne 0}$.
Let $M_t$ be the monomial matrix which is the product of the diagonal matrix
with diagonal elements $\lambda^{t q_i}$ for $i=0,1,\cdots,n-1$ and
the permutation matrix corresponding to the permutation
$i\mapsto t_i$ for $i=0,1,\cdots,n-1$. Then Eq.~\eqref{psi_s} implies that:

$\bullet$\hangindent10mm~
{\em When $\sum_{i=0}^{n-1}a_iX^i$ is viewed as the word
$(a_0,a_1,\cdots,a_{n-1})$,  the map $\M_t$ corresponds
to the monomial equivalence on $F_q^n$ by multiplying 
the monomial matrix $M_t$.}

\noindent
In particular, $\M_t$ is an algebra isomorphism, 
and ${\rm w}(\M_s(a(X)))={\rm w}(a(X))$
for any $a(X)\in R_{n,\lambda}$.
\qed

\medskip
We call $\M_t$ defined in the above lemma 
an {\em isometry} from $R_{n,\lambda}$ to $R_{n,\lambda^{t}}$.
Note that more general isometries were introduced in \cite{CFLL}, but
Lemma \ref{isometry} contains more precise information for our later citations.

For example, if $t=-1$, then we can take $\bar t=nr-1$.
Noting that $\lambda X^{n}\equiv 1\!\pmod{X^n-\lambda^{-1}}$,
for $a(X)=a_0+a_1X+\cdots+a_{n-1}X^{n-1}
%\sum_{i=0}^{n-1}a_iX^i
\in R_{n,\lambda}$ we have
$$
\M_{-1}(a(X))\equiv\lambda X^{n}\sum_{i=0}^{n-1}a_iX^{i(nr-1)}
\equiv\sum_{i=0}^{n-1}\lambda a_iX^{nri}X^{n-i}\pmod{X^n-\lambda^{-1}}.
$$
Since $X^{nr}\equiv 1\!\pmod{X^n-\lambda^{-1}}$, we get 
\begin{equation}\label{M -1}
\M_{-1}\big( a_0+a_{1}X+\cdots+a_{n-1}X^{n-1}\big)
 = a_0+\lambda a_{n-1}X+\cdots+\lambda a_1X^{n-1}.
\end{equation}

\smallskip Next, we refine the set $P_{n,\lambda^t}$ in \eqref{X^n-lambda^t}
and the group $G_{n,r}$ in \eqref{G_n,r} for any $t\in\Z_{nr}^*$.
The decomposition $n=n_rn_r'$ introduced in the following remark 
will be used throughout the paper.

\begin{Remark}\label{n_r'}\rm
Let $n_r'$ be the maximal divisor of the integer $n$ which is coprime to $r$.
Hence $n=n_rn_r'$ such that $n_r'$ is coprime to $r$, 
and $p|r$ for any prime divisor $p|n_r$. 
Let $t\in\Z_{nr}^*$ as before.
By the Chinese Remainder Theorem (cf. \cite[eq.(IV.3), eq.(IV.4)]{CDFL} for details),
\begin{equation}\label{CRT}
\begin{array}{rcl}
P_{n,\lambda^t}=t+r\Z_{nr}&\CRT & (t+r\Z_{n_r r})\times\Z_{n_r'},\\[3pt]
G_{n,r}=\Z_{nr}^*\cap(1+r\Z_{nr})
  &\CRT& (1+r\Z_{n_r r})\times \Z_{n_r'}^*,
\end{array}
\end{equation}
where $\CRT$ stands for the equivalence by the Chinese Remainder Theorem,
and $1+r\Z_{n_r r}$ is a subgroup of $\Z_{n_r r}^*$ with
order $|1+r\Z_{n_r r}|=n_r$ (see Lemma \ref{group n_r} below for more details).
The group  $G_{n,r}$ acts on $P_{n,\lambda^t}$ with $1+r\Z_{n_r r}$ and
$\Z_{n_r'}^*$ respectively acting on $t+r\Z_{n_r r}$ and $\Z_{n_r'}$
respectively.

There is a distinguished subset $P_{n,\lambda^{t}}^{(0)}$
of $P_{n,\lambda^t}$ as follows:
\begin{equation}\label{P^0}
 P_{n,\lambda^{t}}^{(0)}\CRT
  (t+r\Z_{n_r r})\times \{0\}\subseteq (t+r\Z_{n_r r})\times \Z_{n_r'};
\end{equation}
that is, $P_{n,\lambda^{t}}^{(0)}$ consists of
the elements of $P_{n,\lambda^t}$ which are divisible by $n_r'$.
It is easy to see that $P_{n,\lambda^t}^{(0)}$ is $\mu_s$-invariant
for any $s\in G_{n,r}$. In particular,
$P_{n,\lambda^t}^{(0)}$ is a union of some $q$-cosets.
Let $\bar n'_r$ be an integer such that $n'_r\bar n'_r\equiv 1~({\rm mod}~r)$.
For $i\in P_{n,\lambda^t}^{(0)}$, we can write $i=i'n'_r$;
since $i'n'_r\equiv t~({\rm mod}~r)$,
we see that $i'\equiv t\bar n'_r~({\rm mod}~r)$. Thus
$$
 f_{P_{n,\lambda^t}^{(0)}}(X)=
 \prod_{i\in P_{n,\lambda^t}^{(0)}}(X-\theta^i)
  =X^{n_r}-\lambda^{t\bar n'_r}.
$$
\end{Remark}

Generalizing the notations for negacyclic codes in \cite{Bl08},
we make the following definition for general constacyclic codes.

\begin{Definition}\label{def duadic}\rm
Let $t\in\Z_{nr}^*$ and $s\in G_{n,r}$
(then $\M_s$ is an isometry of $R_{n,\lambda^t}$ to itself).
By $C_{n,\lambda^t}^{(0)}=C_{P_{n,\lambda^{t}}^{(0)}}$
we denote the $\lambda^t$-constacyclic code with check set
$P_{n,\lambda^{t}}^{(0)}$, i.e., $X^{n_r}-\lambda^{t\bar n'_r}$
is a check polynomial of $C_{n,\lambda^t}^{(0)}$,
where $\bar n'_r$ is an integer such that $n'_r\bar n'_r\equiv 1~({\rm mod}~r)$.
Let $C\subseteq R_{n,\lambda^t}$ be a $\lambda^t$-constacyclic code.
\begin{itemize}
\item[(i)]
If $R_{n,\lambda^t}= C\oplus \M_s(C)$, i.e.,
$R_{n,\lambda^t}= C+\M_s(C)$ and $C\cap\M_s(C)=0$,
then we say that $C$ and $\M_s(C)$ are a pair of
{\em Type-I duadic $\lambda^t$-constacyclic codes}.
\item[(ii)]
If $R_{n,\lambda^t}=C_{n,\lambda^t}^{(0)}\oplus C\oplus \M_s(C)$,
then we say that $C$ and $\M_s(C)$ are a pair of
{\em even-like duadic $\lambda^t$-constacyclic codes},
which are also named 
a pair of {\em Type-II} duadic $\lambda^t$-constacyclic codes.
\item[(iii)]
If $R_{n,\lambda^t}= C+\M_s(C)$
and $C\cap\M_s(C)=C_{n,\lambda^t}^{(0)}$,
then we say that $C$ and $\M_s(C)$ are a pair of
{\em odd-like duadic $\lambda^t$-constacyclic codes}.
\end{itemize}
\end{Definition}

Note that, if $C$ and $\M_s(C)$ are a pair of even-like (i.e., Type-II)
duadic $\lambda^t$-constacyclic codes,
i.e., $R_{n,\lambda^t}=C_{n,\lambda^t}^{(0)}\oplus C\oplus \M_s(C)$,
then $\M_s^2(C)=C$ because 
$\M_s^2(C_{n,\lambda^t}^{(0)})=C_{n,\lambda^t}^{(0)}$.
It is the same for Type-I duadic $\lambda^t$-constacyclic codes
and odd-like duadic $\lambda^t$-constacyclic codes.

If $C$ and $\M_{s}(C)$ are a pair of even-like 
$\lambda^t$-constacyclic codes, then from the direct sum 
$R_{n,\lambda^t}= C_{n,\lambda^t}^{(0)}\oplus C\oplus \M_{s}(C)$ 
it is easy to see that $C'=C_{n,\lambda^t}^{(0)}\oplus C$ and
$\M_s(C')= C_{n,\lambda^t}^{(0)}\oplus \M_s(C)$ 
are a pair of odd-like duadic $\lambda^t$-constacyclic codes.
 
Conversely, assume that $C'$ and $\M_s(C')$ are a pair 
of odd-like duadic $\lambda^t$-constacyclic codes. 
Since $R_{n,\lambda^t}$ is a semisimple algebra
(i.e. $X^n-\lambda^t$ has no multiple roots), 
there is a unique ideal $C$ such that $C'=C_{n,\lambda^t}^{(0)}\oplus C$, 
hence $\M_s(C')=C_{n,\lambda^t}^{(0)}\oplus \M_s(C)$.
Then $C$ and $\M_s(C)$ are a pair of 
even-like duadic $\lambda^t$-constacyclic codes.

We show an example.
Let $q=5$, $n=6$ and $\lambda=2\in F_5^*$ (so $r=4$). Then
$R_{n,\lambda}=F_5[X]/\langle X^6-2\rangle$, $nr=24$, $n_r=2$, $n_r'=3$,
$P_{n,\lambda}=\{1,5,9,13,17,21\}$, 
$P_{n,\lambda}^{(0)}=\{9,21\}$. 
Take $s=\bar s=13$, then $s\bar s\equiv 1\!\pmod{24}$.
It is easy to check that
$$
X^6-2=(X^2-3)(X^2+X+2)(X^2-X+2),
$$
and $f_{P_{n,\lambda}^{(0)}}(X)=X^2-3$.
Let $C\subseteq R_{n,\lambda}$ be the $\lambda$-constacyclic 
code with check polynomial $X^2+X+2$. 
Since  $\M_{13}(X^2+X+2)=X^2-X+2$, we have the direct sum
$R_{n,\lambda}=C_{n,\lambda}^{(0)}\oplus C\oplus \M_{13}(C)$.
Thus, $C$ and $\M_{13}(C)$ are a pair of 
even-like (Type-II) duadic $\lambda$-constacyclic codes.
On the other hand,
$C_{n,\lambda}^{(0)}\oplus C$ and 
$C_{n,\lambda}^{(0)}\oplus \M_{13}(C)$ are a pair of 
odd-like duadic $\lambda$-constacyclic codes.
In fact, this example is a specific instance of Proposition \ref{a GRS} below.

\smallskip
The following lemma shows that the isometry
$\M_t: R_{n,\lambda}\to R_{n,\lambda^t}$
is related closely to the bijection
$\mu_{t}: P_{n,\lambda}\to P_{n,\lambda^t}$.

\begin{Lemma}\label{bar t P} Let $t\in\Z_{nr}^*$.
If  $C_P\subseteq R_{n,\lambda}$ is a $\lambda$-constacyclic code
with check set $P\subseteq P_{n,\lambda}$,
then
\begin{equation*}
 \M_t(C_P)=C_{tP}~\subseteq~ R_{n,\lambda^t}
\end{equation*}
is a $\lambda^t$-constacyclic code,
i.e., $f_{tP}(X)=\prod_{Q\in tP/\mu_q}f_Q(X)$
is a check polynomial of the $\lambda^{t}$-constacyclic code $\M_t(C_P)$.
\end{Lemma}

\pf Let $c(X)\in C_{P}$. By Remark~\ref{C_P}, $c(\theta^i)=0$
for all $i\in P_{n,\lambda}\backslash{P}$.
By the definition of $\M_t$ in Lemma~\ref{isometry},
$\M _t(c(X))=c(X^{\bar t})+b(X)(X^n-\lambda^t)$ for some $b(X)\in F_q[X]$,
where $t,\bar t\in\Z_{nr}^*$ satisfying that $t\bar t=1~({\rm mod}~nr)$.
Hence
$$
\M _t(c(\theta^i))=c(\theta^{\bar t i})+b(\theta^i)(\theta^{in}-\lambda^t)=0,
\qquad \forall~~i\in {P_{n,\lambda^t}}\backslash{tP}.
$$
So $\M _t(c(X))\in C_{tP}$, see Remark~\ref{C_P}. Since $\M_t$ is an algebra isomorphism,
$$\dim\big(\M_t(C_P)\big)=\dim(C_P)=|P|=|tP|=\dim(C_{tP}),$$
where $|P|$ denotes the cardinality of the set $P$. Thus $\M_t(C_P)=C_{tP}$.
\qed

\medskip
By replacing a permutation equivalence $\Phi$ defined in \cite{Bl13}
with the isometry $\M_{-1}$ defined in Lemma \ref{isometry},
we modify \cite[Th.4]{Bl13} as follows: 

\begin{Lemma}\label{dual code}
Let $C\subseteq R_{n,\lambda}$ be a $\lambda$-constacyclic code. Set
$$
 {\rm Ann}(C)=\big\{a(X)\in R_{n,\lambda}\,\big|\,
   a(X)c(X)=0~{\rm in}~R_{n,\lambda},~\forall~c(X)\in C\},
$$
which is also a $\lambda$-constacyclic code.
Then
$$C^{\bot}=\M _{-1}\big({\rm Ann}(C)\big)~\subseteq~ R_{n,\lambda^{-1}}$$
is a $\lambda^{-1}$-constacyclic code.
\end{Lemma}

\pf
Let $a(X)=\sum_{i=0}^{n-1}a_iX^i\in {\rm Ann}(C)$. In $R_{n,\lambda}$,
since $XX^{n-1}=\lambda$ is invertible, $X$ is invertible.
For $c(X)\in C$,
there is a $b(X)=\sum_{i=0}^{n-1}b_iX^i\in C$ such that $Xb(X)=c(X)$.
In $R_{n,\lambda}$, since $c(X) a(X)=0$, $b(X) a(X)=0$.
Considering the coefficient of $X^{n-1}$, we get
$$ b_0a_{n-1}+b_1a_{n-2}+\cdots+b_{n-1}a_0=0.$$
In $R_{n,\lambda^{-1}}$, by Eq. \eqref{M -1}, 
$$\lambda^{-1}\M_{-1}(a(X))=\lambda^{-1}a_0+ a_{n-1}X+\cdots+a_1X^{n-1}.$$
Noting that, in $R_{n,\lambda}$, $Xb(X)$ is corresponding to the word
$(\lambda b_{n-1},b_0,\cdots,b_{n-2})$, we obtain that
$$\lambda^{-1}\big\langle c(X),\, \M_{-1}(a(X))\big\rangle
 =\big\langle Xb(X), \,\lambda^{-1}\M_{-1}(a(X))\big\rangle=0.
$$
In conclusion, $\M_{-1}(a(X))\in C^\bot$. Thus
$
\M _{-1}\big({\rm Ann}(C)\big)\subseteq C^{\bot}.~
$
Since $\dim C^\bot=n-\dim C=\dim{\rm Ann}(C)$,
we get $C^{\bot}=\M _{-1}\big({\rm Ann}(C)\big)$.
\qed

Since $R_{n,\lambda}$ is a semisimple algebra,
for any $\mu_q$-invariant subset $P\subseteq P_{n,\lambda}$,
it is easy (cf. Eq. \eqref{X^n-lambda}) to see that
\begin{equation}\label{C_{overline P}}
 {\rm Ann}(C_P)=C_{\overline P}, \qquad
 \mbox{where}~~\overline P =P_{n,\lambda}\backslash P.
\end{equation}
Combining it with Lemma \ref{dual code} and Lemma \ref{bar t P},
we have an immediate corollary.

\begin{Corollary}\label{dual code'} With notations in \eqref{C_{overline P}},~
$C_P^\bot =C_{-\overline P}=C_{\overline{-P}}$,~ where
$-P=(-1)P$ and $\overline{-P}=P_{n,\lambda^{-1}}\backslash(-P)$.
\end{Corollary}

\begin{Theorem}\label{even odd}
Let $C\subseteq R_{n,\lambda}$ be a 
$\lambda$-constacyclic code and $s\in G_{n,r}$. 
Then $C$ and $\M_s(C)$ are a pair of even-like duadic 
$\lambda$-constacyclic codes if and only if 
$C^\bot$ and $\M_s(C)^\bot$ are a pair of odd-like duadic
$\lambda^{-1}$-constacyclic codes.
\end{Theorem}

\pf Let $C=C_P$ with check set $P\subseteq P_{n,\lambda}$.
Then $\M_s(C)=C_{sP}$ with check set $sP\subseteq P_{n,\lambda}$.
Assume that $C$ and $\M_s(C)$ are even-like duadic $\lambda$-constacyclic codes,
i.e., $R_{n,\lambda}=C_{n,\lambda}^{(0)}+C_P+C_{sP}$,
$C_{n,\lambda}^{(0)}\cap(C_P+C_{sP})=0$ and $C_P\cap C_{sP}=0$.
By Eq.~\eqref{check set},
$P_{n,\lambda}=P_{n,\lambda}^{(0)}\cup P\cup sP$ and
$P_{n,\lambda}^{(0)}$, $P$, $sP$ are pairwise disjoint.
Note that $\mu_{-1}$ transforms $P_{n,\lambda}$ to $P_{n,\lambda^{-1}}$
bijectively and $-P_{n,\lambda}^{(0)}=P_{n,\lambda^{-1}}^{(0)}$ obviously.
We get that $P_{n,\lambda^{-1}}=P_{n,\lambda^{-1}}^{(0)}\cup(-P)\cup(-sP)$
and $P_{n,\lambda^{-1}}^{(0)}$, $-P$, $-sP$ are pairwise disjoint. So
\begin{eqnarray*}
\overline{-P}&=&P_{n,\lambda^{-1}}\backslash(-P)=
P_{n,\lambda^{-1}}^{(0)}\cup(-sP),\\
\overline{-sP}&=&P_{n,\lambda^{-1}}\backslash(-sP)=
P_{n,\lambda^{-1}}^{(0)}\cup(-P).
\end{eqnarray*}
In $R_{n,\lambda^{-1}}$,  by Corollary \ref{dual code'} we have
$C_P^{\bot}=C_{\overline{-P}}$ and $C_{sP}^\bot=C_{\overline{-sP}}$.
By Eq.~\eqref{check set}, from the above equalities we obtain that
$$
R_{n,\lambda^{-1}}=C_P^{\bot}+C_{sP}^\bot,\quad
C_P^{\bot}\cap C_{sP}^\bot=C_{P_{n,\lambda^{-1}}^{(0)}}
=C_{n,\lambda^{-1}}^{(0)}.
$$
Thus, $C_P^{\bot}$ and $C_{sP}^\bot=\M_s(C_P)^\bot$ are odd-like duadic
$\lambda^{-1}$-constacyclic codes.

Conversely, assume that $C_P^{\bot}$ and $C_{sP}^\bot$ are odd-like duadic
$\lambda^{-1}$-constacyclic codes. It is easy to check that
all the arguments in the above paragraph can be reversed.
Thus we can backward step by step to reach the conclusion that
$C_P$ and $\M_s(C_P)$ are even-like duadic $\lambda$-constacyclic codes.
\qed\smallskip

In the special case where $r=1$ (i.e., cyclic codes are considered) and $s=-1$,
the result \cite[Th. 6.4.2]{HP} is a consequence of the above theorem.

\begin{Lemma}\label{s-iso}
Let $t\in\Z_{nr}^*$, and
$C_P\subseteq R_{n,\lambda}$ be a $\lambda$-constacyclic code
with check set $P\subseteq P_{n,\lambda}$.
Then the following conditions are equivalent.
\begin{itemize}
\item[(i)]
$\M_t(C_P)\subseteq C_P^\bot$ (in the case we call $C$ a
 {\em $\M_t$-isometrically orthogonal code}).
\item[(ii)]
$\M_{-t}(C)\subseteq R_{n,\lambda}$ and $C_P\cap\M_{-t}(C_P)=0$.
\item[(iii)]
$-t\in G_{n,r}$ and $P\cap (-tP)=\emptyset$.
\end{itemize}
\end{Lemma}

\pf (i) $\Leftrightarrow$ (iii).~
By Lemma \ref{bar t P} and Corollary \ref{dual code'},
(i) holds if and only if $C_{tP}\subseteq C_{\overline{-P}}$
where $\overline{-P}=P_{n.\lambda^{-1}}\backslash(-P)$;
by Eq.~\eqref{check set},
it is equivalent to that $tP\subseteq \overline{-P}=-\overline P$
where $\overline{P}=P_{n.\lambda}\backslash P$;
i.e., $-tP\subseteq \overline P$, (iii) holds.

(ii) $\Leftrightarrow$ (iii).~
``$\M_{-t}(C)\subseteq R_{n,\lambda}$'' is obviously equivalent to
``$-t\in G_{n,r}$''.
Note that $\M_{-t}(C_P)=C_{-tP}$. Then, by Eq.~\eqref{check set},
$C_P\cap\M_{-t}(C_P)=0$ if and only if $P\cap (-tP)=\emptyset$.
\qed

\medskip
Taking $t=1$ in Lemma \ref{s-iso}, we get a known consequence:

\begin{Corollary}
A $\lambda$-constacyclic code $C_P\le R_{n,\lambda}$ with check set
$P\subseteq P_{n,\lambda}$ is self-orthogonal if and only if
$\lambda=\pm1$ and $P\cap(-P)=\emptyset$.
\end{Corollary}

Generalizing the self-orthogonality, 
we consider the iso-orthogonality.

\begin{Definition}\label{def-iso-orthog}\rm
Let $C\subseteq R_{n,\lambda}$ be a $\lambda$-constacyclic code.
\begin{itemize}
\item[(i)]
If there is an $s\in G_{n,r}$ such that
$C$ is $\M_{-s}$-isometrically orthogonal (i.e., Lemma \ref{s-iso}(i) for $t=-s$ holds),
then we say that $C$ is
{\em isometrically self-orthogonal}, or {\em iso-orthogonal} for short.
\item[(ii)]
If there is an $s\in G_{n,r}$ such that both $C$ and $\M_s(C)$
are $\M_{-s}$-isometrically orthogonal
(hence $C\cap\M_s(C)=0$, see Lemma \ref{s-iso}(ii))
and $\M_s^2(C)=C$, then we say that $C,\M_s(C)$ are an
{\em iso-orthogonal pair} of $\lambda$-constacyclic codes.
\item[(iii)]
An iso-orthogonal pair $C,\M_s(C)$ of $\lambda$-constacyclic codes
is said to be {\em maximal} if for any
iso-orthogonal pair $C',\M_{s'}(C')$ of $\lambda$-constacyclic codes
we have $\dim C'\le\dim C$.
\end{itemize}
\end{Definition}

If $C_P$, $\M_s(C_P)$ are Type-I duadic $\lambda$-constacyclic codes, i.e.,
$P_{n,\lambda}=P\cup(sP)$ is a partition, %and $P\cap(sP)=\emptyset$,
then $C_P,\M_s(C_P)$ are of course a maximal iso-orthogonal pair
of $\lambda$-constacyclic codes. In fact, in that case
both $C_P$ and $\M_s(C_P)$ are {\em iso-dual $\lambda$-constacyclic codes},
see \cite{Bl13}. Otherwise, if the Type-I duadic constacyclic codes do not exist,
then we show that any pair of even-like duadic constacyclic codes is a
maximal iso-orthogonal pair of constacyclic codes provided it does exist.

\begin{Lemma}\label{type i}
Type-I duadic $\lambda$-constacyclic codes of length $n$ over $F_q$
exist if and only if the order of the quotient group
$(1+r\Z_{n_r r})/\langle q\rangle_{\Z_{n_r r}^*}$ is even,
where $\langle q\rangle_{\Z_{n_r r}^*}$ denotes the subgroup of $\Z_{n_r r}^*$
generated by $q$.
\end{Lemma}

\pf Note that $q\in 1+r\Z_{n_r r}$ hence
$\langle q\rangle_{\Z_{n_r r}^*}\subseteq 1+r\Z_{n_r r}$.
This lemma has been included in \cite[Th.4]{CDFL}
where more complicated results are proved. For convenience,
we sketch a quick proof of the lemma. If
$(1+r\Z_{n_r r})/\langle q\rangle_{\Z_{n_r r}^*}$ is of even order,
we take $s_0\in 1+r\Z_{n_r r}$ such that in the quotient group
the element $s_0$ has order $2$; and take $s\in G_{n,r}$ such that
$s\CRT(s_0,1)\in(1+r\Z_{n_r r})\times\Z_{n_r'}^*$, cf. Eq.~\eqref{CRT}.
By Remark \ref{group theory}(iii) and (iv), it is easy to see that
any $s$-orbit on $P_{n,\lambda}/\mu_q$ has length $2$.
By Remark \ref{group theory}(ii), there is a $\mu_q$-invariant
subset $P\subseteq P_{n,\lambda}$ such that
$P_{n,\lambda}=P\cup sP$ is a partition. %and $P\cap sP=\emptyset$.
Hence $C_P$ and $C_{sP}$ are a pair of Type-I duadic $\lambda$-constacyclic codes.

Conversely, if $C_P$ and $C_{sP}$ are a pair of
Type-I duadic $\lambda$-constacyclic codes,
by Remark \ref{group theory}(ii), the length of any $s$-orbit on
$P_{n,\lambda}^{(0)}/\mu_q$ is even; so in the quotient group
$(1+r\Z_{n_r r})/\langle q\rangle_{\Z_{n_r r}^*}$
the order of the element $s$ is even (cf. Remark \ref{group theory}(i)).
Hence the order of the group
$(1+r\Z_{n_r r})/\langle q\rangle_{\Z_{n_r r}^*}$ is even.
\qed

\begin{Theorem}\label{iso-orth}
Assume that Type-I duadic $\lambda$-constacyclic codes of length $n$ 
do not exist but Type-II (i.e., even-like) duadic $\lambda$-constacyclic 
codes of length $n$ exist. Then any pair $C_P$, $\M_s(C_P)$ of 
Type-II duadic $\lambda$-constacyclic codes of length $n$ 
is a maximal iso-orthogonal pair of $\lambda$-constacyclic codes.
 \end{Theorem}

\pf By Lemma \ref{type i} and the assumption of the theorem,
the order of the quotient group
$(1+r\Z_{n_r r})/\langle q\rangle_{\Z_{n_r r}^*}$ is odd.
Then, by Remark \ref{group theory}(iv), for any $s'\in G_{n,r}$,
the length of any $s'$-orbit on the quotient set
$P_{n,\lambda}^{(0)}\big/\mu_q$ is odd.

Now we prove the theorem by contradiction. Suppose that
$C_{P'}$ and $\M_{s'}(C_{P'})$ are an iso-orthogonal pair
of $\lambda$-constacyclic codes such that
$$\dim C_{P'}>\dim C_P=|P|=\frac{n-n_r}{2}.$$
Set $P''=P'\cap P_{n,\lambda}^{(0)}\subseteq P_{n,\lambda}^{(0)}$.
Since $s'P_{n,\lambda}^{(0)}=P_{n,\lambda}^{(0)}$,
$s'P''=s'P'\cap P_{n,\lambda}^{(0)}\subseteq P_{n,\lambda}^{(0)}$.
Because $P'\cap(s'P')=\emptyset$, we have $P''\cap s'P''=\emptyset$ and
$$
  |P'\cup(s'P')|=|P'|+|s'P'|=2|P'|=2\dim C_{P'}
   >n-n_r=\big|P_{n,\lambda}\backslash P_{n,\lambda}^{(0)}\big|.
$$
Thus, $P''$ and $s'P''$ are non-empty subsets 
of $P_{n,\lambda}^{(0)}$ 
such that $P''\cap s'P''=\emptyset$ and $s'^2P''=P''$.
Note that both $P''$ and $s'P''$ are $\mu_q$-invariant.
The permutation $\mu_{s'}$ gives a bijection from the quotient set
$P''/\mu_q$ to the quotient set $s'P''/\mu_q$.
Thus, the length of any $s'$-orbit on the quotient set
$(P''\cup s'P'')/\mu_q$ is even. This is a contradiction.
\qed

\section{Existence of Type-II duadic constacyclic codes}

We keep notations introduced in Section 2,
and describe the decomposition $n=n_rn_r'$ in Remark \ref{n_r'} more precisely.
Assume that $r_1,\cdots,r_h$, $r'_1,\cdots,r'_{h'}$, $p_1,\cdots,p_\ell$
are distinct primes such that
\begin{equation}\label{n,r}
\begin{array}{l}
 r=r_1^{e_1}\cdots r_h^{e_h}{r'_1}^{e_1'}\cdots{r'_{h'}}^{e'_{h'}};\qquad
 h,h'\ge 0;~~\mbox{all $e_i,e_i'$ are positive;}\\[6pt]
n=r_1^{u_1}\cdots r_h^{u_h}{p_1}^{v_1}\cdots{p_{\ell}}^{v_{\ell}},\qquad
 \ell\ge 1,~~\mbox{all $u_i,v_i$ are positive.}
\end{array}
\end{equation}
Then $n=n_rn_r'$ where
\begin{equation}\label{n_r}
  n_r=r_1^{u_1}\cdots r_h^{u_h},\qquad
  n_r'={p_1}^{v_1}\cdots {p_{\ell}}^{v_{\ell}}.
\end{equation}
In this section we consider Eq.~\eqref{CRT} and Eq.~\eqref{P^0} only for $t=1$,
as restated below.
\begin{equation}\label{CRT'}
\begin{array}{rcl}
P_{n,\lambda}=1+r\Z_{nr}&\CRT & (1+r\Z_{n_r r})\times\Z_{n_r'},\\[3pt]
G_{n,r}=\Z_{nr}^*\cap(1+r\Z_{nr})
  &\CRT& (1+r\Z_{n_r r})\times \Z_{n_r'}^*,
\end{array}
\end{equation}
\begin{equation}\label{P^0'}
 P_{n,\lambda}^{(0)}\CRT (1+r\Z_{n_r r})\times \{0\}\subseteq (1+r\Z_{n_r r})\times \Z_{n_r'}.
\end{equation}

The main result of this section is as follows.

\begin{Theorem}\label{existence}
Type-II duadic $\lambda$-constacyclic codes of length $n$ over $F_q$
exist if and only if one of the following two holds.

(i)~ $n_r$ is even (equivalently, both $n$ and $r$ are even).\nopagebreak

(ii)~ $n$ is odd and $q$ is a square of an element in $\Z_{n_r'}$.
\end{Theorem}

We will prove it in two cases. Case 1: $n_r$ is even, see Theorem \ref{r even} below.
Case~2: $n_r$ is odd, see Theorem~\ref{r odd} below.

Corresponding to Definition \ref{def duadic}, we have the following definition.

\begin{Definition}\label{def type ii}\rm
Let notations be as in Eq.~\eqref{CRT'} and Eq.~\eqref{P^0'}.
Let $P\subseteq P_{n,\lambda}$ and $s\in G_{n,r}$.
\begin{itemize}
\item[(i)]
If $P_{n,\lambda}=P\cup(sP)$ is a partition,
%and $P\cap(sP)=\emptyset$,
then $P$, $sP$ are called a {\em Type-I duadic splitting} of $P_{n,\lambda}$
given by $\mu_s$ (and
$C_P$, $C_{sP}$ are a pair of Type-I duadic $\lambda$-constacyclic codes).
\item[(ii)]
If $P_{n,\lambda}=P_{n,\lambda}^{(0)}\cup P\cup(sP)$
is a partition,
%and $P_{n,\lambda}^{(0)},P,sP$ are disjoint from each other,
then $P$, $sP$ are called a {\em Type-II duadic splitting} of $P_{n,\lambda}$
given by $\mu_s$ (and  $C_P$, $C_{sP}$ are a pair of
Type-II (i.e., even-like) duadic $\lambda$-constacyclic codes in
Definition~\ref{def duadic}(ii)).
\end{itemize}
\end{Definition}

Similarly as in Definition \ref{def duadic},
it is easy to check that with above definitions we have $s^2 P=P$.
%for the (i) and (ii) of the definition,
%it is easy to check that $s^2 P=P$.

We need more precise information on the subgroup $1+r\Z_{n_r r}$ of $\Z_{n_r r}^*$.

\begin{Lemma}\label{group n_r}
With notations in \eqref{n_r}-\eqref{P^0'}, the following hold.
\begin{itemize}
\item[(i)]
$1+r\Z_{n_r r}\CRT (1+r_1^{e_1}\Z_{r_1^{e_1+u_1}})\times\cdots
\times(1+r_h^{e_h}\Z_{r_h^{e_h+u_h}})$,~
and the order of the direct factor
$\big|(1+r_i^{e_i}\Z_{r_i^{e_i+u_i}})\big|=r_i^{u_i}$
for $i=1,\cdots,h$. Hence, the cardinality
$\big|1+r\Z_{n_r r}\big|=\big|P_{n,\lambda}^{(0)}\big|=n_r$.
\item[(ii)]
The group $1+r\Z_{n_r r}$ has even order if and only if
both $n$ and $r$ are even. If this is the case, assuming that
$r_1=2$, $e=e_1\ge 1$ and $u=u_1\ge 1$, we~have
$$1+r\Z_{n_r r}\CRT (1+2^{e}\Z_{2^{e+u}})\times
(1+r_2^{e_2}\Z_{r_2^{e_2+u_2}})\times\cdots
\times(1+r_h^{e_h}\Z_{r_h^{e_h+u_h}})$$
with $1+2^{e}\Z_{2^{e+u}}$ being the Sylow $2$-subgroup of $1+r\Z_{n_r r}$.
\item[(iii)]
Type-I duadic splittings of $P_{n,\lambda}$ exist if and only if
both $n$ and $r$ are even and
 $\langle q\rangle_{\Z_{2^{e+u}}^*}\lneqq
   1+2^e\Z_{2^{e+u}}$, where $\langle q\rangle_{\Z_{2^{e+u}}^*}$
denotes the subgroup of $\Z_{2^{e+u}}^*$ generated by $q$.
\end{itemize}
\end{Lemma}

\pf (i). With the notation in \eqref{n,r}, by the Chinese Remainder Theorem
we have (cf. \cite[eq.(IV.3)]{CDFL} for more details):
\begin{eqnarray*}
1+r\Z_{n_r r}&\CRT& (1+r_1^{e_1}\Z_{r_1^{e_1+u_1}})\times\cdots
\times(1+r_h^{e_h}\Z_{r_h^{e_h+u_h}})\\
&&\quad\times (1+{r'_1}^{e'_1}\Z_{{r'_1}^{e'_1}})\times\cdots
\times(1+{r'_{h'}}^{e_{h'}}\Z_{{r'_{h'}}^{e'_{h'}}}).
\end{eqnarray*}
But $1+{r'_i}^{e'_i}\Z_{{r'_i}^{e'_i}}=\{1\}$ for $i=1,\cdots,h'$,
and $\big|(1+r_i^{e_i}\Z_{r_i^{e_i+u_i}})\big|=r_i^{u_i}$ for $i=1,\cdots,h$.
So (i) holds.

(ii) follows from (i).

(iii).~ By Lemma \ref{type i},
Type-I duadic splittings of $P_{n,\lambda}$ exist if and only if
$(1+r\Z_{n_r r})/\langle q\rangle_{\Z_{n_r r}^*}$ is a group of even order;
 by (ii), this happens if and only if the quotient of the Sylow $2$-subgroup
 $(1+2^e\Z_{2^{e+u}})/\langle q\rangle_{\Z_{2^{e+u}}^*}$
 is non-trivial, which happens if and only if (iii) holds.
\qed

\begin{Theorem}\label{r even} If both $n$ and $r$ are even, then the
Type-II duadic $\lambda$-constacyclic codes of length $n$ over $F_q$ exist.
\end{Theorem}

\pf Let notations be as in \eqref{n_r}-\eqref{P^0'}.
Since both $n$ and $r$ are even,
we can assume, as in Lemma \ref{group n_r}(ii),
that $r_1=2$, $e=e_1\ge 1$ and $u=u_1\ge 1$.
If $P\cup sP$ is a Type-I splitting of $P_{n,\lambda}$ given by $\mu_s$,
then $P'=P\backslash(P_{n,\lambda}^{(0)}\cap P)$ is non-empty and 
it is easy to check that $P'\cup sP'$ is a 
Type-II splitting of $P_{n,\lambda}$ given by $\mu_s$.
So we can further assume that
Type-I duadic $\lambda$-constacyclic codes of length~$n$ do not exist.
Hence, by Lemma \ref{group n_r}(iii),
$q$ generates the multiplicative group $1+2^e\Z_{2^{e+u}}$, i.e.,
${\rm ord}_{\Z_{2^{e+u}}^*}(q)=2^u$.

To prove the existence of Type-II duadic $\lambda$-constacyclic codes of length $n$,
by Remark \ref{group theory} (i), (ii) and (iv),
it is enough to show that there is an integer $s\in G_{n,r}$
such that ${\rm ord}_{\Z_{nr}^*}(s)=2^f$ with $f\ge 1$ and
\begin{equation}\label{sQ neq Q}
sQ\ne Q,\qquad\mbox{ for any $q$-coset $Q$ on~}
 P_{n,\lambda}\backslash P_{n,\lambda}^{(0)}.
\end{equation}

By Lemma \ref{group n_r}(ii), we write
$1+r\Z_{n_r r}=(1+2^e\Z_{2^{e+u}})\times L$ with $L$ being a group of odd order.
By Eq.~\eqref{n_r},  we refine Eq.~\eqref{CRT'} as follows:
\begin{equation*}
\begin{array}{rcl}
P_{n,\lambda}&\CRT & (1+2^e\Z_{2^{e+u}})\times L\times
  \Z_{p_1^{v_1}}\times\cdots\times\Z_{p_\ell^{v_\ell}},\\[3pt]
G_{n,r}  &\CRT& (1+2^e\Z_{2^{e+u}})\times L
 \times \Z_{p_1^{v_1}}^*\times\cdots\times\Z_{p_\ell^{v_\ell}}^*.
\end{array}
\end{equation*}
Let $1\le i\le \ell$. Then $p_i$ is an odd prime and
$\Z_{p_i^{v_i}}^*$ is a cyclic group of order $p_i^{v_i-1}(p_i-1)$.
Since $p_i-1$ is coprime to $p_i^{v_i-1}$,
there is a unique subgroup $H_i$ of $\Z_{p_i^{v_i}}^*$ such that
\begin{equation}\label{H_i}
\Z_{p_i^{v_i}}^*=(1+p_i\Z_{p_i^{v_i}})\times H_i
\end{equation}
and the natural homomorphism $\Z_{p_i^{v_i}}^*\to\Z_{p_i}^*$ induces
an isomorphism $H_i\cong\Z_{p_i}^*$.
Note that $2\,\big|\,(p_i-1)=|\Z_{p_i}^*|$.
We choose an integer $s_i\in\Z_{p_i^{v_i}}^*$ for different cases.

\hangindent54pt Case 1:~
If ${\rm ord}_{\Z_{p_i^{v_i}}^*}(q)$ is odd, we take $s_i\in H_i$ such that
${\rm ord}_{H_i}(s_i)=2^{f_i}$ with $f_i=1$;

\hangindent54pt Case 2:~
if ${\rm ord}_{\Z_{p_i^{v_i}}^*}(q)$ is even, then there is an odd integer $d_i$
such that the order ${\rm ord}_{\Z_{p_i^{v_i}}^*}(q^{d_i})=2^{f_i}$
with $f_i\ge 1$ (hence $q^{d_i}\in H_i$);
in that case, we take $s_i=q^{d_i}$.

\noindent
Let
$$s=(1,1,s_1,\cdots,s_\ell)\in (1+2^e\Z_{2^{e+u}})\times L
 \times \Z_{p_1^{v_1}}^*\times\cdots\times\Z_{p_\ell^{v_\ell}}^*.$$
Then ${\rm ord}_{\Z_{nr}^*}(s)=2^f$ where
$f=\max\{f_1,\cdots,f_\ell\}\ge 1$.

Let $Q$ be any $q$-coset on $P_{n,\lambda}$ outside $P_{n,\lambda}^{(0)}$,
i.e., $Q\subseteq P_{n,\lambda}\backslash P_{n,\lambda}^{(0)}$.
Take
$$(\alpha,\alpha',\alpha_1,\cdots,\alpha_\ell)\in Q~~\mbox{with}~
 \alpha\in 1+2^e\Z_{2^{e+u}},~\alpha'\!\in\! L,~
 \alpha_i\!\in\!\Z_{p_i^{v_i}}~\mbox{for}~i\!=\!1,\cdots,\ell.
$$
Since $q$ generates $1+2^e\Z_{2^{e+u}}$,
we have $q^t\alpha\equiv 1~({\rm mod}~2^{e+u})$ for some integer~$t$.
Set $k'=q^t\alpha'\in L$ and $k_i=q^t\alpha_i\in\Z_{p_i^{v_i}}$
for $i=1,\cdots,\ell$. Then
$$
 (1,k',k_1,\cdots,k_\ell)=q^t(\alpha,\alpha',\alpha_1,\cdots,\alpha_\ell)\in Q.
$$
Now we prove Eq.~\eqref{sQ neq Q} by contradiction.
Suppose that $sQ=Q$. Because $Q\cap P_{n,\lambda}^{(0)}=\emptyset$,
there is an integer $m$ with $1\le m\le \ell$
such that $k_m\not\equiv 0~({\rm mod}~p_m^{v_m})$.
Since $sQ=Q$,
$$ s(1,k',k_1,\cdots,k_m,\cdots,k_\ell)
=(1,k',s_1k_1,\cdots,s_mk_m,\cdots,s_\ell k_\ell)\in Q.$$
Thus, there is an integer $j$ such that
$$
 q^j(1,k',k_1,\cdots,k_m,\cdots,k_\ell)=(1,k',s_1k_1,\cdots,s_mk_m,\cdots,s_\ell k_\ell).
$$
In particular,
$$q^j\equiv 1\pmod{2^{e+u}}\qquad{\rm and}
 \qquad q^jk_m\equiv s_mk_m\pmod {p_m^{v_m}}.$$
Since ${\rm ord}_{\Z_{2^{e+u}}^*}(q)=2^u$, from the first equality
we have $j\equiv 0~({\rm mod}~2^u)$; in particular, $j$ is even.
Next, write $k_m=p_m^{v_m'}k_m'$ with $p_m\nmid k_m'$,
then $0\le v_m'<v_m$ because $k_m\not\equiv 0~({\rm mod}~p_m^{v_m})$.
The second equality becomes:
$$q^jp_m^{v_m'}k_m'\equiv s_mp_m^{v_m'}k_m'~({\rm mod}~p_m^{v_m}).$$
Hence $q^j\equiv s_m~({\rm mod}~p_m^{v_m-v_m'})$. Since $v_m-v_m'\ge 1$, we get
\begin{equation}\label{q^j s_m}
 q^j\equiv s_m \pmod {p_m}.
\end{equation}

In Case 1, in the group $\Z_{p_m}^*$ the order of the element $q^j$ is odd,
but the order of $s_m$ is~$2$; it is a contradiction to Eq.~\eqref{q^j s_m}.

In Case 2, in the group $\Z_{p_m}^*$ the order of the element $q$ is even;
but $j$ is even and $d_m$ is odd, hence
$$\nu_2\big({\rm ord}_{\Z_{p_m}^*}(q^j)\big)
  <\nu_2\big({\rm ord}_{\Z_{p_m}^*}(q)\big)=
  \nu_2\big({\rm ord}_{\Z_{p_m}^*}(q^{d_m})\big),
$$
where $\nu_2(t)$ denotes the $2$-adic valuation of the integer $t$, i.e.
$2^{\nu_2(t)}$ is the maximal power of $2$ dividing $t$.
In particular, $q^j\not\equiv q^{d_m}~({\rm mod}~{p_m})$, which
contradicts Eq.~\eqref{q^j s_m}, as we have chosen $s_m=q^{d_m}$
in this case.

The contradictions finish the proof of the theorem.
\qed

\medskip
Taking $r=2$, from Theorem \ref{r even}
we get the following immediate consequence
which has been proved in \cite{Bl08}.

\begin{Corollary}[\cite{Bl08}] If $n$ is even, then
Type-II duadic negacyclic codes of length $n$ over $F_q$ exist.
\end{Corollary}

By $\nu_2(t)$ we denote the $2$-adic valuation of the integer $t$ as before.

\begin{Theorem}\label{r odd}
Let $n=n_rn_r'$ and $n_r'=p_1^{v_1}\cdots p_{\ell}^{v_\ell}$ 
as in \eqref{n_r}.
Assume that $n_r$ is odd (equivalently, $n$ or $r$ is odd).
Then the following three are equivalent to each other.
\begin{itemize}
\item[(i)]
Type-II duadic $\lambda$-constacyclic codes of length $n$ over $F_q$ exist.
\item[(ii)]
 For all $i=1,\cdots,\ell$, $p_i$ is odd and
 $\nu_2\big({\rm ord}_{\Z_{p_i}^*}(q)\big)<\nu_2(p_i-1)$
 (i.e. $q$ does not generate the Sylow $2$-subgroup of $\Z_{p_i}^*$).
\item[(iii)]
  $n_r'$ is odd and $q$ is a square of an element in $\Z_{n_r'}$.
\end{itemize}
\end{Theorem}

\pf
We refine Eq.~\eqref{CRT'} as follows:
\begin{equation}\label{CRT''}
\begin{array}{rcl}
P_{n,\lambda}&\CRT & (1+r\Z_{n_r r})\times
  \Z_{p_1^{v_1}}\times\cdots\times\Z_{p_\ell^{v_\ell}},\\[3pt]
G_{n,r}  &\CRT& (1+r\Z_{n_r r})
 \times \Z_{p_1^{v_1}}^*\times\cdots\times\Z_{p_\ell^{v_\ell}}^*.
\end{array}
\end{equation}

(i)$\Rightarrow$(ii).
Let $s$ be a multiplier of a Type-II duadic splitting 
$P_{n,\lambda}=P_{n,\lambda}^{(0)}\cup P\cup sP$.
By Lemma \ref{group n_r}(i),
$\big|1+r\Z_{n_r r}\big|=\big|P_{n,\lambda}^{(0)}\big|=n_r$.
So 
$$
 n-n_r=\big|P_{n,\lambda}\backslash P_{n,\lambda}^{(0)}\big|
 =|P|+|sP|=2|P|,
$$
which is an even integer.
%The existence of Type-II splittings of $P_{n,\lambda}$ implies that the cardinality
%$\big|P_{n,\lambda}\backslash P_{n,\lambda}^{(0)}\big| =n-n_r$ is even.
By the assumption of the theorem, $n_r$ is odd. Thus $n$ is odd,
hence $n_r'$ is odd. That is, $p_i$ for $i=1,\cdots,\ell$ are all odd.

Suppose that for some $i$ the inequality in (ii) does not hold, 
without loss of generality, assume that $p_1$ is odd 
and $\nu_2\big({\rm ord}_{\Z_{p_1}^*}(q)\big)=\nu_2(p_1-1)$.
By Eq.~\eqref{CRT''}, we write 
$$
s\CRT (s_0,s_1,s_2\cdots,s_\ell)\in
 (1+r\Z_{n_r r})\times\Z_{p_1^{v_1}}^*\times\Z_{p_2^{v_2}}^*
 \times\cdots\times\Z_{p_\ell^{v_\ell}}^*.
$$
We assume that $S$ is the Sylow $2$-subgroup of $\Z_{p_1^{v_1}}^*$.
Then $|S|=2^{\nu_2(p_1-1)}$,  $q$ generates $S$, 
$\Z_{p_1^{v_1}}^*=S'\times S$ for a subgroup $S'$ of odd order,
and
$$
 (1+r\Z_{n_r r})\times\Z_{p_1^{v_1}}^*
 =(1+r\Z_{n_r r})\times S'\times S
$$
with $(1+r\Z_{n_r r})\times S'$ being a direct factor of odd order.
Thus, the quotient group
\begin{equation*}
\big((1+r\Z_{n_r r})\times\Z_{p_1^{v_1}}^*\big)\big/
\langle q\rangle_{(1+r\Z_{n_r r})\times\Z_{p_1^{v_1}}^*}
\end{equation*}
is of odd order, where
$\langle q\rangle_{(1+r\Z_{n_r r})\times\Z_{p_1^{v_1}}^*}$ is the
subgroup of $(1+r\Z_{n_r r})\times\Z_{p_1^{v_1}}^*$ generated by $q$.
Hence, in the quotient group $(s_0,s_1)$ is an element of odd order.
Take
$$
\alpha\CRT (1,1,0\cdots,0)\in
 (1+r\Z_{n_r r})\times\Z_{p_1^{v_1}}\times\Z_{p_2^{v_2}}
 \times\cdots\times\Z_{p_\ell^{v_\ell}}.
$$
Let $Q_\alpha$ be the $q$-coset containing $\alpha$.
Then $\alpha\notin P_{n,\lambda}^{(0)}$ and
$Q_\alpha\subseteq P_{n,\lambda}\backslash P_{n,\lambda}^{(0)}$.
By Remark \ref{group theory}(iv), the length of the $s$-orbit
on $P_{n,\lambda}/\mu_q$ containing $Q_\alpha$ is odd.
Hence, by Remark \ref{group theory}(ii),
the Type-II splittings of $P_{n,\lambda}$ given by $\mu_s$ do not exist. 
This is impossible because we have had a 
Type-II duadic splitting $P_{n,\lambda}=P_{n,\lambda}\cup P\cup sP$
given by $\mu_s$.  So the equality
$\nu_2\big({\rm ord}_{\Z_{p_1}^*}(q)\big)=\nu_2(p_1-1)$
has to be false.

(ii)$\Rightarrow$(i).~
Assume that (ii) holds. By Eq.~\eqref{H_i},
$\Z_{p_i^{v_i}}^*=(1+p_i\Z_{p_i^{v_i}})\times H_i$ and
the natural homomorphism
\begin{equation*}%\label{natural}
\Z_{p_i^{v_i}}^*~\to~\Z_{p_i}^*,~~~~ 
  k\!\!\!\pmod{p_i^{v_i}}~\mapsto~ k\!\!\!\pmod{p_i}
\end{equation*} 
induces an isomorphism $H_i\cong\Z_{p_i}^*$.
The order of the kernel of the homomorphism is
$|1+p_i\Z_{p_i^{v_i}}|=p_i^{v_i-1}$, 
which is odd and coprime to the order $|\Z_{p_i}^*|=p_i-1$. So
\begin{equation}\label{p^v p}
 \nu_2\big({\rm ord}_{\Z_{p_i^{v_i}}^*}(t)\big)=
 \nu_2\big({\rm ord}_{\Z_{p_i}^*}(t)\big),~~~\mbox{for any integer $t$ coprime to $p$.}
\end{equation}
Then $\nu_2\big({\rm ord}_{\Z_{p_i^{v_i}}^*}(q)\big)=
 \nu_2\big({\rm ord}_{\Z_{p_i}^*}(q)\big)<\nu_2(p_i-1)
 =\nu_2\big(|\Z_{p_i^{v_i}}^*|\big)$.
Since $\Z_{p_i^{v_i}}^*$ is a cyclic group,
%there is a cyclic subgroup $K_i$ of $\Z_{p_i^{v_i}}^*$ such that
%$K_i\supset  \langle q\rangle_{\Z_{p_i^{v_i}}^*}$ and the index
%$\big|K_i:\langle q\rangle_{\Z_{p_i^{v_i}}^*}\big|=2$,
%where $\langle q\rangle_{\Z_{p_i^{v_i}}^*}$ stands for the subgroup
%of $\Z_{p_i^{v_i}}^*$ generated by $q$. So 
there is an $s_i\in\Z_{p_i^{v_i}}^*$ such that 
in the group $\Z_{p_i^{v_i}}^*$ we have
%$s_i\notin\langle q\rangle_{\Z_{p_i^{v_i}}^*}$ but
$s_i^2 = q$ and
$\nu_2\big({\rm ord}_{\Z_{p_i^{v_i}}^*}(s_i)\big)=
 \nu_2\big({\rm ord}_{\Z_{p_i^{v_i}}^*}(q)\big)+1$.
By \eqref{p^v p} again, we obtain that
$\nu_2\big({\rm ord}_{\Z_{p_i}^*}(s_i)\big)=
 \nu_2\big({\rm ord}_{\Z_{p_i}^*}(q)\big)+1$.
So we get:
\begin{equation}\label{s_i}
 s_i\notin\langle q\rangle_{\Z_{p_i}^*}
  \quad\mbox{but}\quad
 s_i^2 \equiv q\pmod{p_i^{v_i}},\qquad i=1,\cdots,\ell.
\end{equation}
By Eq.~\eqref{CRT''} we have an integer $s$ as follows:
$$
 s\CRT (1,s_1,\cdots,s_\ell)\in (1+r\Z_{n_r r})
 \times \Z_{p_1^{v_1}}^*\times\cdots\times\Z_{p_\ell^{v_\ell}}^*.
$$
By Eq.~\eqref{s_i},
in the quotient group $G_{n,r}/\langle q\rangle_{G_{n,r}}$, the element
$s$ has order $2$, where $\langle q\rangle_{G_{n,r}}$ stands for the subgroup
of $G_{n,r}$ generated by $q$.

Let $Q$ be any $q$-coset on $P_{n,\lambda}$ outside $P_{n,\lambda}^{(0)}$.
We prove by contradiction that $sQ\ne Q$,
which implies that the length of
the $\mu_s$-orbit on $P_{n,\lambda}/\mu_q$ containing $Q$ is even
(see Remark \ref{group theory} (i) and (iv)),
hence the statement (i) of the theorem holds, see Remark \ref{group theory} (ii).
Suppose that $sQ=Q$. Take any
${\bf k}=(k_0,k_1,\cdots,k_\ell)\in Q$ with
$$k_0\in 1+r\Z_{n_r r}, \qquad
 k_i\in\Z_{p_i^{v_i}}\quad\forall~ i=1,\cdots,\ell.
$$
Then there is an integer $d$ such that $s{\bf k}=q^d{\bf k}$.
Since $Q\cap P_{n,\lambda}^{(0)}=\emptyset$,
there is an $m$ with $1\le m\le \ell$ such that
$k_m\not\equiv 0~({\rm mod}~p_m^{v_m})$.
But $sk_m\equiv q^d k_m~({\rm mod}~p_m^{v_m})$.
By the argument for Eq.~\eqref{q^j s_m}, 
we have $s\equiv q^d~({\rm mod}~p_m)$,
which implies that $s_m\in\langle q\rangle_{\Z_{p_m}^*}$.
That is a contradiction to Eq.~\eqref{s_i}.

(ii)$\Rightarrow$(iii).~ Taking $s'\in\Z_{n_r'}^*$ such that
$s'\CRT(s_1,\cdots,s_\ell)\in
 \Z_{p_1^{v_1}}^*\times\cdots\times\Z_{p_\ell^{v_\ell}}^*$
where $s_i$ for $i=1,\cdots,\ell$ are taken in Eq.~\eqref{s_i},
we obtain $s'^2\equiv q\pmod{n_r'}$.

(iii)$\Rightarrow$(ii). Assume that $s'^2\equiv q\pmod{n_r'}$.
For $p_i$ with $i=1,\cdots,\ell$, we have $s'^2\equiv q\pmod{p_i}$.
If ${\rm ord}_{\Z_{p_i}^*}(s')$ is odd, then
${\rm ord}_{\Z_{p_i}^*}(q)$ is odd, hence
$\nu_2\big({\rm ord}_{\Z_{p_i}^*}(q)\big)=0<\nu_2(p_i-1)$.
Otherwise, ${\rm ord}_{\Z_{p_i}^*}(s')$ is even, hence
$$\nu_2\big({\rm ord}_{\Z_{p_i}^*}(q)\big)
 =\nu_2\big({\rm ord}_{\Z_{p_i}^*}(s'^2)\big)
 <\nu_2\big({\rm ord}_{\Z_{p_i}^*}(s')\big)\le\nu_2(p_i-1).
$$
The proof of the theorem is finished. \qed

\smallskip
In the extreme case where $n_r=1$ (whether $r=1$ or not),
$n=n_r'$ and Eq.~\eqref{CRT'} becomes
\begin{equation*}
\begin{array}{rcl}
P_{n,\lambda}&\CRT & \{1\}\times \Z_{n}\cong \Z_{n},\\[3pt]
G_{n,r}
  &\CRT& \{1\}\times \Z_{n}^*\cong \Z_{n}^*.
\end{array}
\end{equation*}
Thus, we can view even-like duadic cyclic codes as a special case of
Type-II (even-like) duadic constacyclic codes.

\begin{Corollary}[{\cite[Theorem 6.3.2]{HP}}]
Even-like (i.e., Type-II) duadic cyclic codes of length $n$ over $F_q$ exist
if and only if $n$ is odd and $q$ is a square of an element in~$\Z_{n}$.
\end{Corollary}

\section{Examples}

We show that some good codes can be constructed from
Type-II (even-like) duadic constacyclic codes.

%Let $F_{q^t}$ be an extension of $F_q$. Let
%$\mb{\alpha}=(\alpha_0,\alpha_1,\cdots,\alpha_{n-1})
% \in F_{q^t}^n$ with coefficients different from each other,
%and $\mb{\varepsilon}=(\varepsilon_0,\varepsilon_1,\cdots,\varepsilon_{n-1})
%\in F_{q^t}^n$ with coefficients all non-zero. The following
%$[n,k]$ linear code over $F_{q^t}$:
%\begin{align*}
%&{\rm GRS}_{k}(\mb{\alpha};\mb{\varepsilon})=\\
%&\quad \big\{\big(\varepsilon_0 f(\alpha_0),\varepsilon_1 f(\alpha_1),\cdots,
%\varepsilon_{n-1} f(\alpha_{n-1})\big)  ~\big|~ 
% f(X)\in F_{q^t}[X],~\deg f(X)<k\big\}
%\end{align*}
%is called a {\em generalized Reed-Solomon code}, or {\em GRS-code} for short.
%And, the subfield subcode ${\rm GRS}_{k}(\mb{\alpha};\mb{\varepsilon})|_{F_q}$,
%which is the code over $F_q$ by restricting
%the code ${\rm GRS}_{k}(\mb{\alpha};\mb{\varepsilon})$ over $F_{q^t}$
%to the subfield $F_q$, is said to be an {\em alternant code}, cf. \cite[Ch.9]{LX}.
%Obviously, ${\rm GRS}_{k}(\mb{\alpha};\mb{\varepsilon})$
%is an $[n,k,n-k+1]$ MDS-code, and the minimum distance of its alternant code
%is at least $n-k+1$.

\begin{Proposition}\label{a GRS}
Assume that $\nu_2(q-1)\ge 2$
(equivalently, $\nu_2(q+1)=1$).
Let $n=q+1=2n'$, $r=2^{\nu_2(q-1)}$, $r'=\frac{q-1}{r}$
and $s=1+rn'$. Let
$$
 P=\Big\{ 1+ri ~\Big|~ \frac{n'+r'}{2}<i<\frac{3n'+r'}{2} \Big\}
  \subseteq P_{n,\lambda}=1+r\Z_{nr}.
$$
Then $P$ is $\mu_q$-invariant and
$C_P$, $\M_s(C_P)$ are a pair of even-like $\lambda$-constacyclic codes
of length $n$, which are alternant codes over $F_q$
from generalized Reed-Solomon codes over $F_{q^2}$;
in particular, they are $[q+1,\frac{q-1}{2},\frac{q+5}{2}]$ MDS-codes.
\end{Proposition}

\pf Denote $e=\nu_2(q-1)\ge 2$.
Note that $n_r=2$, $n_r'=n'$ is odd and $q=1+rr'=1+2^{e}r'$ with $r'$ being odd.
So $q$ generates the group $1+r\Z_{n_r r}=1+2^e\Z_{2^{e+1}}$.
In particular, Type-I duadic $\lambda$-constacyclic codes of length $n$
over $F_q$ do not exist, cf. Lemma \ref{group n_r} (iii).

Since $0<r'=\frac{q-1}{2^e}<\frac{q+1}{2}=n'$,  we have
\begin{equation*}\textstyle
 r'<\frac{n'+r'}{2}<n'<\frac{n'+r'}{2}+n'=\frac{3n'+r'}{2}<n.
\end{equation*}
And $|P|=n'-1$. Since $\frac{rr'}{2}+1=\frac{q-1}{2}+1=n'$,
$$\textstyle
 1+r\frac{n'+r'}{2} =\frac{r}{2}n'+\frac{rr'}{2}+1 =(\frac{r}{2}+1)n'.
$$
And $\frac{3n'+r'}{2}=\frac{n'+r'}{2}+n'$. By Eq.~\eqref{P^0},
$$\textstyle
 P_{n,\lambda}^{(0)}=\{1+r\frac{n'+r'}{2},~1+r\frac{3n'+r'}{2}\}.
$$
For any $1+ri\in P_{n,\lambda}$, noting that $q=1+rr'=n-1$, we have
\begin{eqnarray*}
q(1+ri)&=&q+qri=1+rr'+nri-ri\\
&\equiv& 1+r(r'-i)\equiv 1+r(n+r'-i) \pmod{nr}.
\end{eqnarray*}
If $\frac{n'+r'}{2}<i<\frac{3n'+r'}{2}$, it is easy to check that
$$\textstyle
 \frac{n'+r'}{2}<n+r'-i<\frac{3n'+r'}{2}.
$$
Thus, the subset $P$ is $\mu_q$-invariant. Next we compute
\begin{eqnarray*}
 s(1+ri)&=&\textstyle (1+rn')(1+ri)=1+r(n'+i)+\frac{r}{2}nri\\
 &\equiv&1+r(n'+i)\pmod{nr}.
\end{eqnarray*}
Since $P$ consists of the points $1+ri$ with $i$ running
from $\frac{n'+r'}{2}+1$ to $\frac{n'+r'}{2}+n'-1$ consecutively,
we obtain that $P\cap sP=\emptyset$.
As $|P_{n,\lambda}|=2n'=|P_{n,\lambda}^{(0)}|+|P|+|sP|$,
we further obtain that $P=P_{n,\lambda}^{(0)}\cup P\cup sP$.
In conclusion, $C_P$ and $\M_s(C_P)$ are a pair of even-like duadic
$\lambda$-constacyclic codes over $F_q$.

Because $nr$ is a divisor of $(q+1)(q-1)=q^2-1$,
in the extension $F_{q^2}$ of $F_q$ we can take
a primitive $nr$-th root $\theta$ of unity such that $\theta^n=\lambda$.
By $\widetilde C_P$, $\M_s(\widetilde C_P)$ we denote the pair of
even-like duadic $\lambda$-constacyclic codes over $F_{q^2}$. It is clear that
$C_P=\widetilde C_P|_{F_q}$ is the subfield subcode from $\widetilde C_P$.
It is the same for $\M_s(C_P)$. 

Let 
$\hat C=\big\{{\bf c}_f\,\big|\, f=f(X)\in F_{q^2}[X], \deg f<n'-1\big\}
\subseteq F_{q^2}^n$ with
$$
{\bf c}_f=\big(f(1),~\theta^{-(rz+1)}f(\theta^{-r}),~ \cdots, ~
  \theta^{-(rz+1)(n-1)} f(\theta^{-r(n-1)})\big)\in F_{q^2}^n,
$$
where $z=\frac{n'+r'}{2}+1$. 
Then $\hat C$ is a generalized Reed-Solomon code.
To complete the proof of the proposition,
it is enough to show that $\widetilde C_P=\hat C$.

Any codeword ${\bf c}_f$ of $\hat C$ for $f(X)=\sum_{k=0}^{n'-2}f_kX^k$
corresponds to the following $F_{q^2}$-polynomial
$$
 c_f(X)=\sum_{j=0}^{n-1} \theta^{-(rz+1)j}f(\theta^{-rj})X^j
 =\sum_{j=0}^{n-1} \theta^{-(rz+1)j}\sum_{k=0}^{n'-2} f_k \theta^{-rjk}X^j.
$$
Let $1+rm\in \overline P=P_{n,\lambda}\backslash P$,
i.e., $0\le m<z$ or $z+n'-1\le m<n$.
Note that $\theta^{-r}$ is a primitive $n$-th root of unity. 
For any $i$ with $z\le i\le z+n'-2$,
we have $\theta^{-r(i-m)}\ne 1$. Then
\begin{eqnarray*}
c_f(\theta^{1+rm})&=&\sum_{j=0}^{n-1}\sum_{k=0}^{n'-2}
\theta^{-(rz+1)j}f_k \theta^{-rjk}\theta^{(1+rm)j}\\
&=&\sum_{k=0}^{n'-2}f_k\sum_{j=0}^{n-1}\theta^{-r(z+k-m)j}\\
&=&\sum_{k=0}^{n'-2}f_k\cdot
  \frac{\theta^{-r(z+k-m)n}-1}{\theta^{-r(z+k-m)}-1}=0.
\end{eqnarray*}
Since $\overline P$ is the defining set of $\widetilde C_P$,
we get that $c_f(X)\in \widetilde C_P$, see Remark \ref{C_P}. Thus
$\hat C\subseteq \widetilde C_P$.
This inclusion has to be an equality because the dimensions of the two hand sides
are equal to each other.  \qed

\medskip
The following is a specific numerical example of Proposition \ref{a GRS}.

\begin{Example}\rm Take $q=13$, $n=14$, $r=4$ (i.e., $\lambda=5$).
Then $n_r=2$,~ $n'=n_r'=7$, $nr=56$ and
$$P_{n,\lambda}=\{1,5,9,13,17,21,25,29,33,37,41,45,49,53\},$$
which is partitioned into $q$-cosets as follows:
\begin{align*}
&Q_0=P_{n,\lambda}^{(0)}=\{21,49\},\\
&Q_1=\{1,~13\},~~~Q_2=\{5,9\},~~~Q_3=\{17,53\},\\
&Q_4=\{25,45\},~~~Q_5=\{29,41\},~~~Q_6=\{33,37\};
\end{align*}
and
$\prod_{i\in P_{n,\lambda}^{(0)}}(X-\theta^i)=X^2+5. $
Take $s=1+rn'=29$, then $s^2\equiv 1\pmod{56}$.
Then $\mu_s$ permutes the quotient set $P_{n,\lambda}/\mu_q$ of $q$-cosets
into four orbits as follows:
$$  (Q_0)(Q_1,Q_5)(Q_2,Q_6)(Q_3,Q_4). $$
Let
$$ P=Q_4\cup Q_5\cup Q_6=\{25,29,33,37,41,45\}; $$
then
$$ sP=Q_1\cup Q_2\cup Q_3=\{1,5,9,13,17,53\}.$$
Obviously,
$$ P_{n,\lambda}^{(0)}\cup P\cup sP=P_{n,\lambda},\qquad
 P_{n,\lambda}^{(0)}\cap P=P_{n,\lambda}^{(0)}\cap sP=P\cap sP=\emptyset. $$
Thus $C_P$, $C_{sP}$ are a pair of even-like duadic $\lambda$-constacyclic codes
over $F_{13}$ with parameters $[14,6,9]$.
\end{Example}

Finally we exhibit an example where $n_r$ is odd.

\begin{Example}\rm Take $q=4$, $n=21$, $r=3$
hence $F_4=\{0,1,\lambda,\lambda^2\}$. Then
$n_r=3$,~ $n_r'=7$, $nr=63$ and
$$P_{n,\lambda}=\{1,4,7,10,13,16,19,22,25,28,31,34,37,40,43,46,49,52,55,58,61\},$$
which is partitioned into $q$-cosets as follows:
\begin{align*}
&Q_0=P_{n,\lambda}^{(0)}=\{7,28,49\},\\
&Q_1=\{1,~4,~16\},~~~Q_2=\{10,40,34\},~~~Q_3=\{13,52,19\},\\
&Q_4=\{22,25,37\},~~~Q_5=\{31,61,55\},~~~Q_6=\{43,46,58\};
\end{align*}
and
$$ \prod_{i\in P_{n,\lambda}^{(0)}}(X-\theta^i)=X^3-\lambda. $$
Since  $5^2\equiv 4=q\pmod 7$, even-like duadic $\lambda$-constacyclic
codes exist. Take $s=55\equiv-8\pmod{63}$, then $s^2\equiv 1\pmod{63}$.
Then $\mu_s$ permutes the quotient set $P_{n,\lambda}/\mu_q$ of $q$-cosets
into four orbits as follows:
$$  (Q_0)(Q_1,Q_5)(Q_2,Q_6)(Q_3,Q_4). $$
Let
$$ P=Q_1\cup Q_2\cup Q_3=\{1,4,10,13,16,19,34,40,52\}; $$
then
$$ sP=Q_4\cup Q_5\cup Q_6=\{22,25,31,37,43,46,55,58,61\}. $$
Obviously,
$$ P_{n,\lambda}^{(0)}\cup P\cup sP=P_{n,\lambda},\qquad
 P_{n,\lambda}^{(0)}\cap P=P_{n,\lambda}^{(0)}\cap sP=P\cap sP=\emptyset. $$
Thus $C_P$, $C_{sP}$ are a pair of even-like duadic $\lambda$-constacyclic codes
over $F_4$ with parameters $[21,9,d]$, where the minimum distance $d\ge 8$
%(in fact $d=8$)
since the defining set of $C_{sP}$ is
$$P_{n,\lambda}\backslash{sP}=P_{n,\lambda}^{(0)}\cup P
 =\{1,4,7,10,13,16,19,28,34,40,49,52\},
$$
which contains $7$ consecutive points $1,4,7,10,13,16,19$ of $P_{n,\lambda}$.
\end{Example}

\section*{Acknowledgements}
The research of the authors is supported
by NSFC with grant number 11271005.
It is the authors' pleasure to thank the anonymous referees for 
their helpful comments which led to improvements of the paper.

\end{document}